\documentstyle[10pt,aas2pp4]{article}

\def\kpc{\hbox{kpc}}
\def\pc{\hbox{pc}}
\def\kms{\hbox{km}\,\hbox{s}^{-1}}
\def\msun{\hbox{M}_\odot}

\righthead{Milky Way Bulge Optical Depth}
\begin{document}

\title{The Maximum Optical Depth Towards Bulge Stars From \\
Axisymmetric Models of the Milky Way}

\author{Konrad Kuijken\altaffilmark{1}}
\affil{Kapteyn Instituut, PO Box 800, 9700 AV Groningen, The Netherlands}
\altaffiltext{1}{Visiting Scientist, Dept. of Theoretical Physics, 
University of the Basque Country}

\begin{abstract}
It has been known that recent microlensing results towards the bulge
imply mass densities that are surprisingly high given dynamical
constraints on the Milky Way mass distribution. We derive the maximum
optical depth towards the bulge that may be generated by axisymmetric
structures in the Milky Way, and show that observations are close to
surpassing these limits. This result argues in favor of a bar as a
source of significantly enhanced microlensing. Several of
the bar models in the literature are discussed.
\end{abstract}

\keywords{Galaxy: kinematics and dynamics --- Galaxy: structure --- 
gravitational lensing}

\section{Introduction}

Large surveys for microlensing events in the Galaxy have grown into a
serious tool for evaluating models of the mass distribution of the
Milky Way, as is evident from the recent results of the MACHO
(\cite{macho97}), OGLE (\cite{ogle94}), EROS (\cite{eros96}) and DUO
(\cite{duo95}) surveys.  Microlensing occurs whenever the light from a
distant star passes sufficiently close to an intervening object (the
`lens') that its path is significantly deflected. The deflection
results in a focusing of the source star's rays, which may be detected
as an apparent brightening of the source with a characteristic light
curve.  Typically the brightening lasts for a number of days or months
before the source and lens proper motions spoil the alignment. Both
the optical depth $\tau$, which is the average fraction of the sources
that is being amplified more than a factor of 1.34 at any given time,
and the average duration of the individual events, depend on the
distribution and masses of lensing objects between the observer and
the sources (\cite{pac86}).

In this paper we concentrate on microlensing towards the bulge of the
Milky Way. Both the MACHO and OGLE teams find that the optical depth
in this direction is rather higher than had been expected initially:
the optical depths reported are in the range of 3--6$\times10^{-6}$,
roughly twice the expected value based on standard Galaxy
models. Although initially this discrepancy was attributed to our
Galaxy's central bar (\cite{pacetal94}, \cite{zsr95}, \cite{zrs96}),
evidence for which had been accumulating for some time
(e.g. \cite{bs91}, \cite{weinberg92}; for a review, see \cite{kk96}),
the most recent models for the bar in fact do little to enhance the
expected optical depth over that expected from axisymmetric models
(\cite{bissantz96}, \cite{nikwein97}). It therefore seems opportune to
return to the axisymmetric mass models once more.

In this letter we derive a strong limit on axisymmetric models: a
given optical depth towards the bulge comes with a minimum
contribution to the Galactic circular speed at the sun. It turns out
that the current values for the optical depth towards the bulge are
already close to implying a circular speed higher than that observed.

The only result from microlensing theory that is relevant for this
discussion is the optical depth $\tau$ to a source at distance $L$,
due to a distribution of lenses with mass density $\rho(D)$ at
distance $D$ down the line of sight:
\begin{equation}
\label{eq:tau}
\tau={4\pi G\over c^2}\int_0^L \rho(D) {D(L-D)\over L} dD.
\end{equation}
$\tau$ is independent of the mass function $f(m)$ of the lenses,
though the rate and detectability of microlensing events are not:
the survey teams correct for detection (in)efficiency in their
analyses.

\section{Can Axisymmetric Mass Components Explain Bulge Microlensing?}

Axisymmetric density distributions can be viewed as constructed from
concentric rings. We therefore start by considering the relation
between the mass and the optical depth of a ring.  We assume that the
sources are on the Galactic minor axis, at a distance of $R_0=8\kpc$,
and not mixed spatially with the ring; these results are thus
appropriate for the microlensing of the bulge clump giants
(\cite{macho97}).

\subsection{Optical Depth from a Ring}

Consider a ring of uniform surface density with radius $R<R_0$ and
radial width $\Delta R\ll R$. Its optical thickness towards bulge sources
near the Galactic minor axis, at latitude $b$ radians (assumed small),
is (cf. eq.~\ref{eq:tau})
\begin{equation}
\tau_{\rm ring}={4\pi G\over c^2}{(R_0-R)R\over R_0}\rho(z)\Delta R.
\end{equation}
Here $\rho$ is the density where the line of sight intersects the ring, 
at height
$z=(R_0-R)b$ from the plane.  Now, if the ring is axi\-symmetric
with uniform surface density $\Sigma$, then its mass is
\begin{equation}
\label{eq:minring}
M_{\rm ring}=2\pi R\Delta R\Sigma=
{c^2\tau_{\rm ring} R_0b\over2G}\left(\Sigma\over z\rho(z)\right)
\end{equation}
(using $R_0-R=z/b$). The bracketed factor is dimensionless,
and depends only on the shape of the vertical density profile of the
ring, not on the total mass of the ring. We will call it the vertical
shape factor $F_z$.  Its denominator tends to zero at small and large $z$
(provided the ring has finite surface mass density), and therefore
this geometrical factor has a minimum at some intermediate
$z$. Hence equation~\ref{eq:minring} can be used to derive a lower
limit on the mass of an axisymmetric ring of a given optical depth
$\tau$.

The lowest vertical shape factor for monotonic $\rho(|z|)$ is obtained
when the density is constant up to a height $a>z$, and zero beyond. In
this case $\Sigma=2a\rho$, and $F_z\geq 2$. More realistic density
profiles yield values considerably larger,
though. Table~\ref{tab:fz} lists the lower limits to $F_z$ for some
different forms of $\rho(z)$.
In light of those results, we adopt $F_z\gtrsim 4.0$.

\begin{table}
\caption{The minimum vertical shape factor $F_z=\Sigma/(z\rho)$ 
for different vertical density profile shapes.}
\label{tab:fz}
\begin{tabular}{ccc}
$\rho(z)$	&	$F_z^{\rm (min)}$	& Attained at $z=$\\
\tableline
constant to $z=a$	&	2	&	$a$\\
$\exp(-|z|/a)$  	&	$2e$	&	$a$\\
$\hbox{sech}(z/a)$ 	&	$4.7$	&	$1.2a$\\
$\hbox{sech}^2(z/a)$ 	&	$4.5$	&	$0.8a$\\
$\exp(-z^2/a^2)$ 	&	$4.1$	&	$0.7a$\\
\end{tabular}
\end{table}

The lower mass limit of Eq.~\ref{eq:minring} can be recast as a
circular velocity at the solar circle: approximating the circular
speed due to a ring by its monopole term (itself a lower limit), we
obtain
\begin{equation}
v_{\rm circ}^2\geq {GM_{\rm ring}\over R_0}=
{c^2\tau_{\rm ring}b\over2}F_z
\gtrsim 2.0 c^2\tau_{\rm ring} b.
\end{equation}

Since the radius of the ring is no longer explicit in this equation,
this limit applies equally to superpositions of rings, i.e.,
axisymmetric mass distributions.  Scaling to observed parameters, we
obtain
\begin{equation}
v_{\rm circ}^2\geq\left(210\kms\right)^2
\left(\tau_{-6}\over4.0\right)
\left(b\over3.5^\circ\right)\left(F_z^{\rm (min)}\over4.0\right)
\end{equation}
as the minimum circular speed generated at the solar position by
axisymmetrically distributed matter responsible for an optical depth
of $\tau_{-6}\times10^{-6}$ towards bulge sources at Galactic latitude~$b$.
A radially extended mass distribution, in order to lens optimally,
needs to have $F_z$ close to its minimum at all radii: this is
achieved when the scale height of the mass is proportional to $R_0-R$,
i.e., tapers linearly to zero at the solar radius.

\subsection{Specific Disk Models}

\begin{figure}
\epsfxsize=\hsize\epsfbox{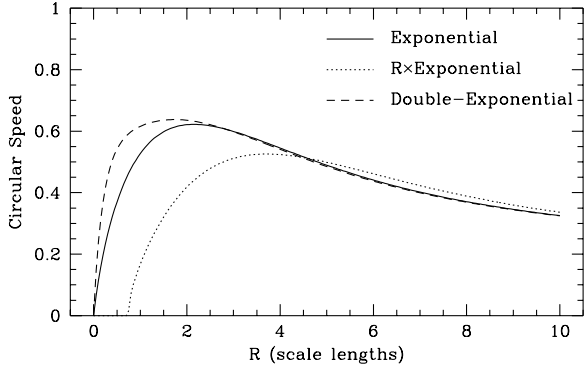}
\caption
{The rotation curves for three types of disk models
considered, in units of $(GM/h)^{1/2}$.
\label{fig:rcs}}
\end{figure}

We know more about the Milky Way rotation curve than its value at the
solar position: we also know that the circular speed does not
greatly exceed the local value at any point in the inner disk of the
Galaxy. We now investigate how this extra constraint affects the
maximum disk optical depth, by considering various models for the disk
mass distribution and raising their mass until their rotation curve
reaches $200\kms$ at any point. We consider four functional forms for
the disk surface mass density, whose rotation curves are shown in
Fig.~\ref{fig:rcs}. We will also compare these models to
solar-neighborhood normalized models, where the
surface mass density in stars is $\sim35\msun\pc^{-2}$ (\cite{kg89b}).
Note that the total surface density of matter within 1 kpc of the plane, 
$71\msun\pc^{-2}$ (\cite{kg89b}) is
consistent with such maximal disk models only for short
($\lesssim2.5\kpc$) scale length models (\cite{sackett97}).

\begin{figure}
\epsfxsize=\hsize\epsfbox{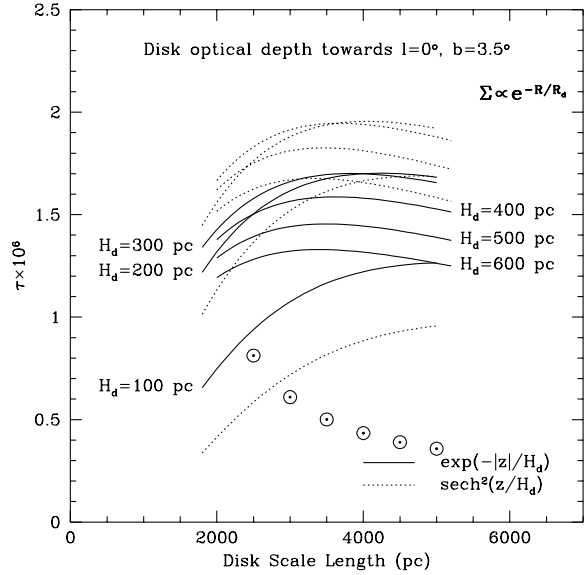}
\caption
{The optical depth towards sources at the Galactic Minor axis
from radially exponential disks with maximum circular speed $200\kms$,
and differing scale heights and lengths. The solid lines refer to
vertically exponential disks, the dotted ones to sech$^2$ disks. Sun
symbols show the optical depths from disks normalized to the `locally
identified' stellar surface density of $35\msun\pc^{-2}$ at the solar
position, vertically exponential with scale height 300pc.
\label{fig:tauexp}}
\end{figure}

The traditional disk model is a radially exponential disk,
$\Sigma(R)\propto \exp(-R/h)$. We consider two vertical density
profile shapes, a vertical exponential and a $\hbox{sech}^2(z/a)$
`isothermal' model. If the scale height is small, the rotation curve
peaks at a radius of 2.2 scale lengths at the value $0.62(GM/h)^{1/2}$
(\cite{freeman70}). Figure~\ref{fig:tauexp} shows the optical depth from
such disks towards the Galactic minor axis at a latitude of
$b=3.5\arcdeg$, for various scale heights and lengths. 

\begin{figure}
\epsfxsize=\hsize\epsfbox{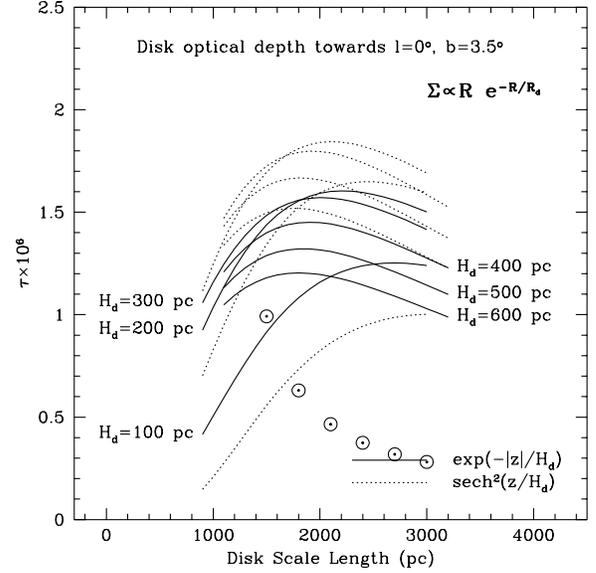}
\caption
{As Figure \ref{fig:tauexp}, but with a central hole in the
disk mass profile.
\label{fig:taurexp}}
\end{figure}

\begin{figure}
\epsfxsize=\hsize\epsfbox{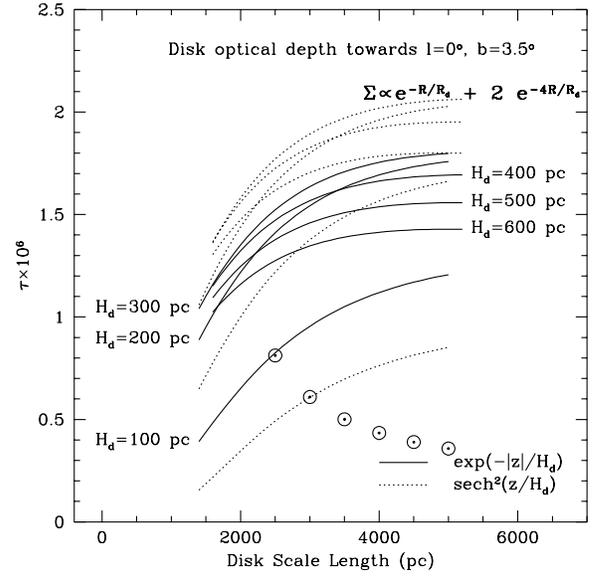}
\caption
{As Figure \ref{fig:tauexp}, 
but with a more concentrated mass
profile.
\label{fig:tautwoexp}}
\end{figure}

For comparison, we consider also two variations on the exponential
disk: a less centrally concentrated model with $\Sigma(R)\propto
R\exp(-R/h)$ (e.g., \cite{caldostr81}), and a more concentrated one
obtained by adding a second exponential disk of a quarter the scale
length, and one eighth the mass of the main component. As may be seen
from Figs.~\ref{fig:taurexp} and~\ref{fig:tautwoexp}, the maximum
optical depth from such disks is comparable to the exponential disk.

Finally, we consider a Mestel disk (\cite{mestel63}), which has
surface mass density $\Sigma(R)=V^2/(2\pi GR)$, and whose circular
speed is everywhere equal to $V$. Such a model microlenses more efficiently
than an exponential disk since it can satisfy our rotation curve
constraint at every radius simultaneously. The maximum
optical depth from a Mestel disk towards $b=3.5\arcdeg$ turns out to
be $2.5\times10^{-6}$ for a vertically sech$^2(z)$ mass
distribution.

We summarize the results for the various disk models in
Table~\ref{tab:disksumm}. We give two numbers for each model: the
maximal optical depth for each model assuming constant scale height,
and the (higher) optical depth that can be achieved by tapering the
disk scale height down to zero at the solar position, in proportion to
$(R_0-R)$.
\begin{table}
\caption{Maximum optical depth towards stars on the Galactic minor
axis at latitude $b=3.5\arcdeg$, for various disk surface density
profiles $\Sigma(R)$. All models have been normalized to have a peak
circular speed of $200\kms$, and have vertical density
$\propto$sech$^2(z)$. The optical depths for models whose scale height
is radially constant, and tapered $\propto(R_0-R)$ are shown.}
\label{tab:disksumm}
\small
\begin{tabular}{lcc}
Disk $\Sigma(R)$&$\tau_{\rm const}\times10^6$&$\tau_{\rm tapered}\times10^6$\\
\tableline
$\propto\exp(-R/a)$			&2.0	&2.6	\\
$\propto R\exp(-R/a)$			&1.9	&2.4	\\
$\propto\exp(-R/a)+2\exp(-R/4a)$	&2.1	&2.7	\\
$\propto R^{-1}$ 			&2.5	&3.3	\\
\end{tabular}
\end{table}

\section{Discussion}

The calculations presented above show that axisymmetric models of the
Galaxy cannot avoid high circular velocities at the sun if the optical
depth towards the Galactic minor axis is as high as is being
reported. The current numbers are in fact so high as to exceed the
maximal disk values derived in Table~\ref{tab:disksumm}. They are also
in conflict with the claimed detection of the dark halo towards the
LMC (\cite{macholmc96}, \cite{eros96}), since the small disk scale
heights ($\sim300\pc$) required for bulge microlensing would produce
virtually no microlensing towards the LMC.  If indeed the optical depth
is too high, then non-axisymmetric models, particularly the Galactic
bar, should be considered.

Several in-depth studies of microlensing by the central bulge/bar have
been carried out already. While the recent investigations by the
Oxford group (\cite{bgs96}; \cite{bissantz96})
and by \cite{nikwein97} conclude that the enhancement in
optical depth over a standard bulge model (\cite{kent92}) is small, the
earlier analysis by Zhao et al. (1995,1996) did find a significant
enhancement in $\tau$. These differences illustrate the difficulty in
disentangling the structure of the inner Galaxy from our view point in
the disk. 

The Zhao et al.\ and Oxford studies are based on the COBE/DIRBE map of
near-infrared emissions, the asymmetries of which are used to infer
the bar's orientation and shape from perspective effects. Zhao et al.\
based their model on the fitting to the DIRBE data by \cite{dwek95},
whose best-fit model is quite elongated, with major axis close to the
line of sight. This model is therefore efficient at microlensing
towards the bulge, since it places most of its mass density between us
and the bulge. The Oxford group, on the other hand, use a Lucy
deprojection of the emission to infer the bar properties. They find a
much shorter bar, which consequently generates less
microlensing. Nikolaev \& Weinberg derive their bar parameters from
star counts of IRAS AGB sources. Their bar is considerably longer than
the COBE-derived ones, but it is quite broad, providing only a modest
enhancement in lensing over axisymmetric models.

It is difficult to prefer one of these models over another. The Zhao
et al.\ model is a simple parametric fit to the COBE data, whereas the
Oxford model is completely non-parametric. In fact, the Oxford
inversion is puzzlingly robust: different starting conditions result
in very similar solutions, in spite of the fact that a
three-dimensional emissivity model is reconstructed out of a
two-dimensional projected surface brightness map! One assumption made
is, however, crucial: the models are taken to be 8-fold symmetric,
i.e., the bar is modeled as reflection symmetric in three orthogonal
planes (it is this symmetry that allows the observed left-right
differences in the map to be interpreted as perspective effects). In
reality, even the most regular bars observed do not exhibit this
symmetry: bars are generally somewhat twisted, and often connect onto
spiral arms joining at the ends. As Binney et al. themselves
recognize, even a 4-fold symmetry (reflection in Galactic plane and
through the Galactic center) is insufficient to allow a stable
inversion.

The Weinberg bar may be stronger than it appears in the analysis of
IRAS sources, since a relatively small number of harmonics was used in
modeling the star counts ($m=0,2,4$ azimuthal harmonics). It is
therefore possible that the true density enhancement towards the bulge
is higher than modeled. This bar is also clearly larger than the
COBE models, suggesting that the Galaxy may well be a bar-within-a-bar
galaxy, a type of system which is relatively common. In that case, the
COBE models may need to take account of this extra foreground
component.

\section{Conclusion}

We have shown that if the mass responsible for the microlensing
towards the Galactic bulge is distributed axisymmetrically, the high
optical depths $\tau$ being reported may be translated into simple but
significant lower limits on the circular speed at the solar position
due to lensing matter:
\begin{equation}
v_c^2\geq 2.0 c^2\tau b
\end{equation}
where $b$ is the galactic latitude (in radians) of the sources. This
limit evaluates to $v_c\geq210\kms$ for an optical depth of
$4\times10^{-6}$ at $b=3.5\arcdeg$, very close to the actual rotation
speed of $200\pm20\kms$ (\cite{merr92}).

This limit is approached when the lensing matter has a scale height
that is well matched to the height of the line of sight towards the
stars being monitored, otherwise even higher circular speeds are
implied. More restricted, but realistic, models for the
mass distribution fail to generate optical depths above 2.5$\times
10^{-6}$ without requiring circular speeds in excess of $200\kms$.

This result comes close to ruling out purely axisymmetric models for
the mass distribution in the inner Galaxy. Those models left require a
very flattened distribution of the Galactic mass, of thickness
$\sim300\pc$, contrary to the conclusions drawn from the microlensing
experiments towards the Large Magellanic Cloud, which favor a round
dark matter distribution. A significant enhancement in $\tau$ from a
central bar appears to be a more plausible explanation.

\bigskip

It is a pleasure to thank Penny Sackett for discussions and a careful
reading of the manuscript.

\end{document}